\documentclass{article}

\usepackage{arxiv}

\usepackage[utf8]{inputenc} 
\usepackage[T1]{fontenc}    
\usepackage{hyperref}       
\usepackage{url}            
\usepackage{booktabs}       
\usepackage{amsfonts}       
\usepackage{nicefrac}       
\usepackage{microtype}      
\usepackage{lipsum}
\usepackage{multicol}
\usepackage{listings}
\usepackage{caption}
\usepackage{multirow}

\usepackage{subfig}
\usepackage{graphicx}

\usepackage{algorithm}
\usepackage{algpseudocode}
\floatname{algorithm}{Algorithm}
\MakeRobust{\Call}

\algdef{SE}[VARIABLES]{Variables}{EndVariables}
   {\algorithmicvariables}
   {\algorithmicend\ \algorithmicvariables}
\algnewcommand{\algorithmicvariables}{\textbf{GPU Related Variables}}

\title{ILP-M Conv: Optimize Convolution Algorithm for Single-Image Convolution Neural Network Inference on Mobile GPUs}

\author{
  Zhuoran Ji \\
  Department of Computer Science\\
  The University of Hong Kong\\
  Hong Kong, China \\
  \texttt{jizr@hku.hk} \\
}

\begin{document}
\maketitle

\begin{abstract}
Convolution neural networks are widely used for mobile applications. However, GPU convolution algorithms are designed for mini-batch neural network training, the single-image convolution neural network inference algorithm on mobile GPUs is not well-studied. After discussing the usage difference and examining the existing convolution algorithms, we proposed the ILP-M convolution algorithm. The ILP-M convolution algorithm achieves $14.6 \times$ speedup than the most popular \textit{im2col} convolution algorithm, and $2.30 \times$ speedup than the fastest existing convolution algorithm (direct convolution) as far as we know. 
\end{abstract}


\section{Introduction}

In recent years, many deep learning applications are meant for edge computing platforms, such as mobile phones and smart IoTs. One notable group of these applications is related to computer vision like object recognition, object tracking, face recognition, and art style transfer. As most of these applications are time-critical, executing them on remote servers and returning the results by the internet has lots of problems due to the internet connection unreliability and delays associated with network latency. Thanks to the advance of mobile system-on-chips (SoC), the computing power of edge computing platforms are enough to execute the convolutional neural network (CNN) inference. There are many attempts to port neural network inference to edge computing platforms, showing the capability of local inference.

Convolution is the fundamental operation of the convolutional neural networks. The computation of batched many-channels convolution is data-intensive and massively parallel, which is naturally executed on Single-Instruction-Multiple-Data (SIMD) processors. Therefore, it is quite popular to use graphics processing units (GPUs) to accelerate the convolutional neural network training. The existing GPU convolution algorithms are designed for mini-batch CNN training, while edge computing platforms usually execute single-image CNN inference. Even both are convolution, there are significant gaps between them, if not entirely different stories, due to the input data size, hardware gaps, and different engineering considerations.

However, few studies have discussed the difference between mini-batch CNNs training on workstations and single-image CNNs inference on edge computing platforms, let alone designing convolution algorithms specialized for the latter. As our evaluation and experiments will show, many widely used algorithms, which achieve excellent performances of mini-batch CNNs training, may perform poorly in single-image CNNs inference on edge computing platforms.

In this paper, we discussed the differences between single-image CNNs inference and mini-batch CNNs training to identify the demand of the single-image CNNs inference. We also analyzed and evaluated some of the most popular GPU convolution algorithms in perspective of single-image CNNs inference. Furthermore, we proposed a novel GPU 2D convolution algorithm specialized for single-image CNNs inference on edge computing platforms, named Instruction-Level Parallelism Maximizing convolution, or ILP-M convolution. ILP-M convolution eliminates the inner-loop memory barrier of direct convolution algorithm by mapping the threads to different output channels rather than pixels. Therefore, ILP-M significantly improves the instruction-level parallelism and reduces registers usage.

The rest of this paper is organized as follows. Section 2 discusses the differences between single-image CNNs inference on edge computing platforms and mini-batch CNNs training on high-end GPUs. Section 3 analyzed and evaluated the existing GPU convolution algorithm. Section 4 proposes the ILP-M convolution algorithm, which is heavily optimized for single-image CNNs inference. Sections 5 describes the experiment results and detailed profile metrics in terms of memory and arithmetic. Section 6 summarizes this paper and suggests future work.

\section{Single-Image CNNs Inference on Embedded GPUs}

Even though both are convolution, single-image CNNs inference on embedded GPUs is quite different from mini-batch CNNs training on high-end dedicated GPUs, if they are not entirely different stories. The differences mainly come from three aspects: the disparity of input images numbers, the hardware gaps between embedded GPUs and dedicated GPUs, and different engineering considerations.

\subsection{Single-Image Limits Threads-Level Parallelism}

\begin{figure}[h!]
\centering

\includegraphics[width=0.45\linewidth]{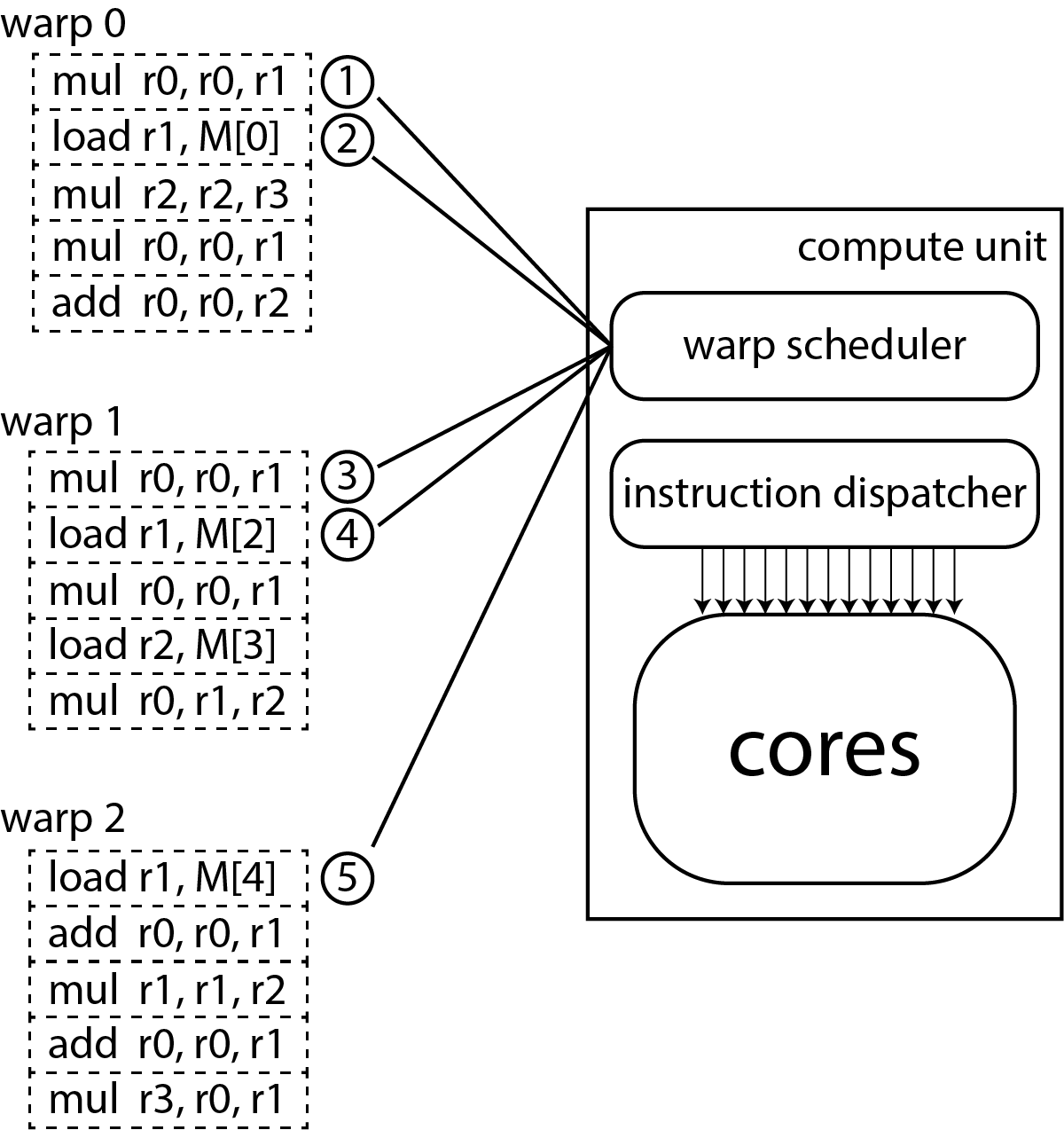}

\caption{Illustration of the thread-level parallelism: the warp scheduler first issues instructions (1) and (2) of warp 0 and be blocked by (2), as it is long-latency global memory access instruction. Instead of waiting for the finish of this instruction, the warp scheduler fetches instructions from the warp 1 to keep the GPU busy. \label{tlp_fig}}

\end{figure}

The most critical difference or challenge of single-image CNNs inference is that only one image is fed into the neural network. For a single image, the insufficiency of the input data prevents us from using as many threads as mini-batch training, which limits the thread-level parallelism. However, thread-level parallelism is the most important mechanism for GPUs to hide the latency. As illustrated in Figure \ref{tlp_fig}, there are many warps in a compute-unit, and the warp scheduler can fetch instructions from any warp independently as long as the warp is not blocked. When a warp issues a  long-latency global memory access instruction, this warp is blocked, and the warp scheduler will fetch instructions from other non-blocked warps rather than wait the global memory access to complete. The thread-level parallelism decreases the arithmetic logic unit from stalling and improves overall GPUs utilization. However, the insufficiency of the warps in a compute-unit limits the compute-unit to take advantage of thread-level parallelism, as the non-blocked warps may be consumed soon.

\begin{figure}[h!]
\centering

\includegraphics[width=0.45\linewidth]{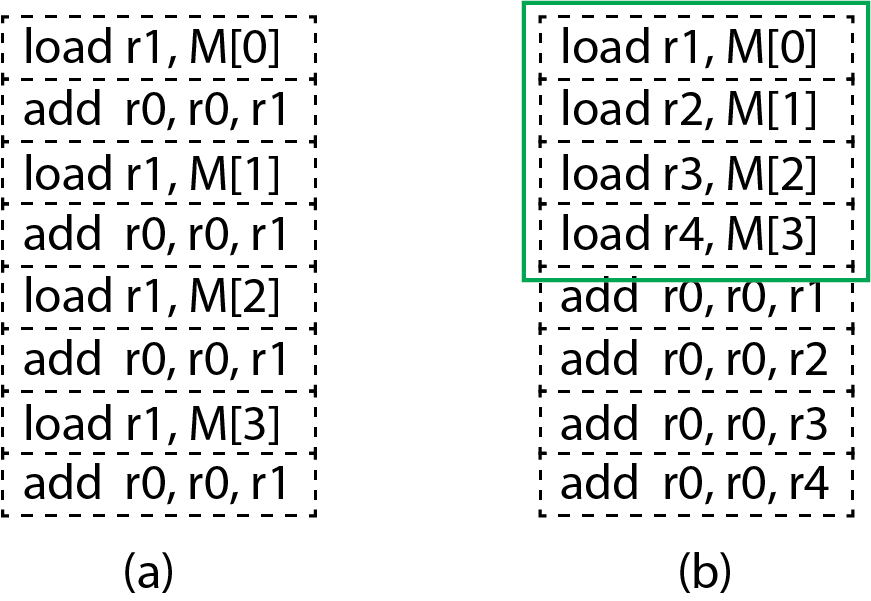}

\caption{Illustration of the instruction-level parallelism \label{ilp_fig}}

\end{figure}

Without enough thread-level parallelism, the single-image CNNs inference needs to seek latency hiding from the other mechanism: instruction-level parallelism. Instruction-level parallelism hides the latency in a single thread by issuing independent instruction simultaneously. If an instruction is independent with its previous instructions, this instruction can be issued no matter whether the previous instructions have finished or not. As shown in Figure \ref{ilp_fig}(b), the first four instructions load four independent values from global memory, which are independent of each other. The second instruction can be issued even if the first instruction has not finished. It usually relies on compilers to use instruction scheduling and register reallocation to reorder the instructions to maximize the instruction-level parallelism. The code in Figure \ref{ilp_fig}(a) does the same work with Figure \ref{ilp_fig}(b) without any instruction-level parallelism. Every addition instruction needs the value loaded by its previous instruction, and every global memory access instruction needs to wait its previous addition instruction to free the register (\texttt{r1}).

However, two common constraints restrict the usage of instruction scheduling to improve instruction-level parallelism. The first constraint is introduced by memory barriers. It is a common optimization strategy to reduce the global memory access by letting threads of a workgroup copy data from global memory to the shared memory collaboratively when each value is used by multiple threads. As a thread needs the values loaded by others, before reading the shared memory, a memory barrier is needed to guarantee every value has been written into the shared memory. The compiler cannot schedule memory instructions across memory barriers, which restricts the instructions reordering and therefore the instruction-level parallelism.

Another constraint is the additional usage of registers. As shown in Figure \ref{ilp_fig}, the (a) only uses two registers, while the (b) needs five registers. The simultaneously loaded values need to be stored in different registers, which increases the registers usage significantly. As most GPUs do not support inter-threads register reuse, the registers need to be allocated to a warp during launching and reserved in the whole lifetime of the warp. If a memory-then-compute group, for example the first two instructions of Figure \ref{ilp_fig}(a), needs a great number of registers, pipelining too many such groups makes the registers become the bottleneck of the resource demands. It may reduce the number of warps that a compute-unit can hold and therefore decreases the thread-level parallelism, which is the primary mechanism for GPUs to hide the latency. As the number of warps is unknown during compilation time, the compiler cannot estimate the thread-level parallelism and may restrict the register usage to balance the thread-level parallelism and instruction-level parallelism. Therefore, the register usage also restricts the instruction-level parallelism.

\subsection{Memory Bandwidth and Energy Consumption of Embedded GPUs}
Another challenge of executing inference on embedded GPUs comes from hardware limitations. Most embedded GPUs and integrated GPUs use LPDDR4 or DDR4 as off-chip global memory, whose bandwidth is far less than GDDR6 and HBM2. The memory bandwidth of dual-channel LPDDR4 is about 30GB/s, while the bandwidth of GDDR6 and HBM2 is about 600GB/s and 1TB/s, respectively. Even worse, the limited memory bandwidth of mobile GPUs is shared by CPUs and other processors of the SoC, leading to even lower real memory bandwidth. It is much more easily for the global memory access to become the bottleneck on mobile GPUs. 

Additionally, the off-chip memory access consumes tens of times the energy compared with on-chip cache access and hundreds of times the energy compared with floating-point arithmetic. Even though energy consumption becomes to draw attentions of deep learning areas, it is still the last consideration when designing GPUs convolution algorithms, especially for GPUs that are powered by mains electricity. However, edge computing platforms are usually battery-powered. The energy consumption determines the battery lifetime. On embedded GPUs, it should be more careful when trading-off between global memory access and other operations.



\subsection{Engineering Choice of Inference}

Last but not least, the engineering choice of the convolution algorithms for CNNs inference is usually different from that of CNNs training. For CNNs training, various neural network architectures and combinations of convolution parameters are examined to find the one with the best performance. In such a scenario, an algorithm that provides stable and acceptable performance for any convolution setting is preferred. It frees the practitioners from tuning the GPU kernels for every convolution parameter, and the practitioners can focus on the performance of the CNNs. However, for the CNNS inference, the neural network architecture and convolution parameter is fixed. In this stage, the goal becomes to optimize the GPUs convolution kernel codes for short inference time and low energy consumption. It is worth to adopt the convolution algorithm that achieves the fastest speed even if great efforts need to be paid for tuning.


\section{Existing GPU Convolution Algorithms}

This section reviews several popular GPU convolution algorithms, which are unrolling-based convolution algorithms(\texttt{im2col} and \texttt{libdnn}), Winograd convolution algorithm, and direct convolution algorithm. The FFT-based convolution algorithm is not discussed. It performs well only with large convolution filters, while the state-of-the-art CNNs mainly use small convolution filters.

\subsection{Unrolling-based Convolution Algorithms}

\begin{figure}[h!]
\centering

\includegraphics[width=0.8\linewidth]{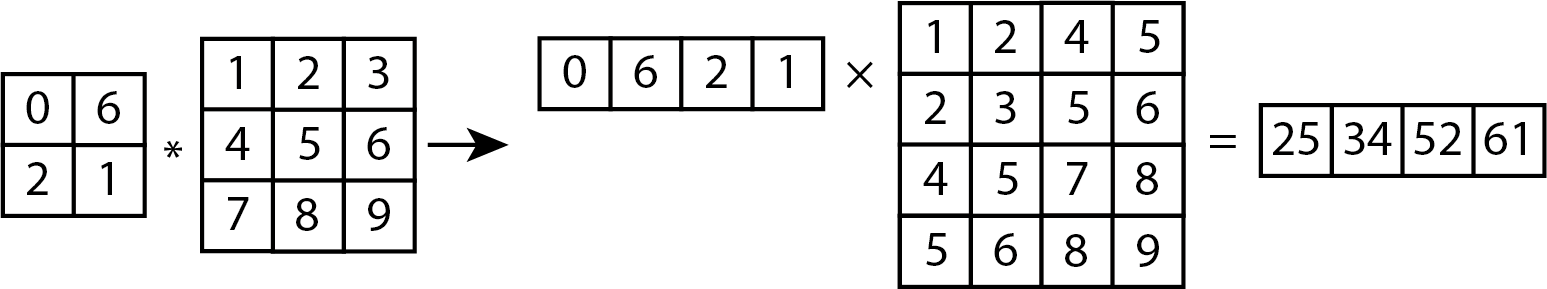}

\caption{Illustration of the \texttt{im2col} convolution algorithm \label{im2col}}

\end{figure}

Unrolling-based convolution algorithm is the most popular GPU convolution algorithm for CNNs training. The key idea is to transform the sliding-window convolution, which is hard to optimize, into well-studied matrix multiplication \cite{im2col}. As shown in Figure \ref{im2col}, each convolution filter is flattened into a row, while the input images are unrolled into columns of a matrix. Each pixel of the output image is then computed by dot-product of each column of the unrolled input matrix with the corresponding convolution filter row. The transformation is named \texttt{im2col}, and we denoted this unrolling-based convolution algorithm as \texttt{im2col} convolution. It allows performing convolution with heavily optimized GEMM libraries, such as clBLAS and cuBLAS.

The limitations of the \texttt{im2col} convolution are that it wastes memory to store duplicated input images and introduces significant global memory access overhead. As the BLAS libraries provide GEMM as a standalone function, \texttt{im2col} convolution separates the \texttt{im2col} and matrix multiplication to two GPU kernels. The \texttt{im2col} GPU kernel needs to store the unrolled input matrix into global memory, and then the GEMM kernel needs to load the unrolled input matrix back from global memory. As the size of the unrolled input matrix is \texttt{kernel\_size} times of the input images, it incurs significant global memory bandwidth overhead, especially for embedded GPUs.

Another unrolling-based convolution implementation named \texttt{libdnn} \cite{tschopp2016efficient, Tschoppgithub} eliminates the global memory bandwidth overhead by combining the \texttt{im2col} and GEMM into a single GPU kernel. When performing the tiled matrix multiplication, each tile of the unrolled input matrix is constructed on the fly only by the workgroups that need this tile. As all tiles of the unrolled input matrix are only stored in on-chip memory and discarded after matrix multiplication, they do not need to be written to and read from the global memory.


Even though \texttt{libdnn} eliminates the storage and bandwidth overhead incurred by the unrolled input matrix, the performance of \texttt{libdnn} convolution is not always better than the \texttt{im2col} convolution, especially for CNNs training. First, as the \texttt{im2col} kernel only performs index calculation and global memory copy, which can be heavily pipelined with thread-level parallelism, considering the state-of-the-art dedicated GPUs have up to 1TB/s global memory bandwidth. Also, as each tile of the unrolled input matrix is used by multiple workgroups, many workgroups need to unroll the same tile. The unrolling operation involves complex index calculation and irregular global memory access, which are unfriendly to GPUs.

\subsection{Winograd Convolution Algorithm}

Winograd convolution algorithm was documented in 1980, but it has not been widely used by CNNs until 2015 \cite{lavin2016fast}. The Winograd convolution algorithm first divides the output images into tiles and computes each tile as follows, $$A^{T}[(Gg) \odot (B^{T}d)]$$ where $g$ is the convolution filters, $d$ is the input images. The $A$, $B$, and $G$ are transformation matrices, which are constants for a given value of tile size and convolution filter size. For Winograd convolution with tile size $M$ and convolution filter size $R$, each tile needs $(M+R-1)^{2} \times R^{2}$ times multiplication, while the unrolling-based convolution and direct convolution need $M^{2}R^{2}$ times multiplication. Meanwhile, the Winograd convolution also increases the thread-level parallelism, as the matrix multiplication of smaller transformed matrices has more independent workload and therefore more warps.

The reduction of the number of multiplication is at the cost of extra global memory access and floating-point addition. As the convolution filters are also constants for CNNs inference, only the transformation of input images and inverse transformation of the output images need to be performed. The number of global memory access and addition increases quadratically with the tile size and convolution filter size, or formally, $$A(transformation\_cost) \in \mathcal{O}(M^{2} + R^{2})$$ For large convolution filters, the memory access and addition complexity of the transformation may overwhelm the benefits of multiplication reduction for large convolution filters. Also, the transformations introduce substantial global memory access, which is expensive for embedded GPUs due to the bandwidth limitation and energy consumption. 



\subsection{Direct Convolution Algorithm}

Direct convolution is the convolution algorithm that follows the definition of convolution \cite{krizhevsky2012cuda}. The convolution filters slide over the input images, and the dot-product between their elements are computed and accumulated into the corresponding output images. Many studies \cite{perrot2016optimized, iandola2013communication, chen2017optimizing, li2015fast} optimized global memory access and data reusing in on-chip memory to speedup the trivial direction convolution algorithm. These optimizations are similar to that of matrix multiplication, which mainly consists of collaboratively loading the data from global memory and assigning more work to a thread to reuse the register. Generally, the direct convolution needs less shared memory compared with other convolution algorithms, as it caches the input images rather than the transformed or unrolled data. Meanwhile, the direct convolution algorithm usually has no complex index and memory offset calculation and therefore needs fewer arithmetic instructions.


\begin{algorithm}
\caption{Direct Convolution Algorithm \label{dcap}}
\begin{algorithmic}[1]

\Function{conv\_cache\_filter}{}

    \For{$1$ to IN\_CHANNELS}
        \State \Call{load}{$img\_global$, $img\_shared$}
        \For{$1$ to OUT\_CHANNELS\_PRE\_THREAD}
            \State \Call{load}{$filter\_global$, $filter\_shared$}
            \State \Call{barrier}{LOCAL\_MEM}
            \State add $filter\_shared \bigodot img\_shared$ into $out\_reg$
        \EndFor 
    \EndFor
    \State \Call{save}{$out\_reg$, $output\_global$}
\EndFunction
\State
\Function{conv\_nocache\_filter}{}

    \For{$1$ to IN\_CHANNELS}
        \State \Call{load}{$img\_global$, $img\_shared$}
        \State \Call{barrier}{LOCAL\_MEM}
        \For{$1$ to OUT\_CHANNELS\_PRE\_THREAD}
            
            \State add $filter\_global \bigodot img\_shared$ into $out\_reg$
        \EndFor
    \EndFor
    \State \Call{save}{$out\_reg$, $output\_global$}
\EndFunction

\end{algorithmic}
\end{algorithm}


However, there are lots of parameters and decisions that affect the performance of the direct convolution, and it usually needs more efforts to tune the direct convolution GPU kernels to find out the optimal combination of parameters. The main computation of unrolling-based convolution and Winograd convolution is matrix multiplication, whose key parameters are only the tile shape and workload per threads. Direct convolution algorithm has all GEMM's parameters and additional parameters that are difficult to choose, such as the output channels processed per threads and whether the convolution filters and outputs should be cached in the shared memory.

Among all implementation choices of direct convolution kernels, the most critical contradiction for single-image CNNs inference is that whether to cache the convolution filters in the shared memory or not. The pseudocode of the caching implementation is shown in Algorithm \ref{dcap} (\texttt{CONV\_CACHE\_FILTER}), where the convolution filters are loaded from global memory and cached in the shared memory collaboratively. Each thread only needs to load a small portion of rather than the whole convolution filters from the global memory, which reduce the global memory bandwidth pressure. However, a memory barrier needs to be put before performing the dot-product to guarantee all weights of the convolution filters have been written into the shared memory. For CNNs training, the latency is hidden by thread-level parallelism, while for CNNs inference, the instruction-level parallelism is the primary mechanism for latency hiding. There are only \texttt{filter\_size} times arithmetic and no memory loading between any two adjacent inner barriers (Line 6). It limits the instructions scheduling significantly, as the compiler cannot move memory instructions across memory barriers. Even worse, as the convolution filters are stored in the same piece of the shared memory, the global memory loading instructions cannot be issued until the previous arithmetic has finished. It entirely prevents the compiler from fusing the memory instructions and arithmetic instructions to hide the latency.

On the other hand, the pseudocode of the non-caching implementation is listed in Algorithm \ref{dcap} (\texttt{CONV\_NOCACHE\_FILTER}). In this case, the dot-product involves filter size times global memory access and \texttt{filter\_size} times arithmetic. There are \texttt{output\_channel\_per\_thread $\times$ filter\_size} times independent arithmetic and memory instructions between two adjacent memory barriers for the compiler to reorder to hide the latency. However, there is a noticeable drawback if the convolution filters are not loaded collaboratively. As each thread needs to load all convolution filters, the non-caching implementation needs much more global memory access. The L2 cache may ease the problem, but the overhead is still substantial. The duplicated global memory access also increases the registers used to store the same values. To load the convolution filters simultaneously, the loaded values need to be stored in different registers. Pipelining the calculation within a dot-product needs \texttt{filter\_size} times registers, which significantly increases the register usage and therefore restricts the instruction-level parallelism. Meanwhile, as the latency of memory instructions are usually longer than that of floating-point arithmetic instructions, the ratio of these two kinds of instructions of direct convolution is unfriendly to instruction-level parallelism.



    

        
        
        
    
        




\section{Methodology} 

In this paper, we proposed a GPU convolution algorithm for single-image CNNs inference on embedded GPUs, named Instruction-Level Parallelism Maximizing convolution (ILP-M convolution). It is based on the direct convolution algorithm but optimized for single-image by eliminating the memory barriers without introducing duplicated convolution filters loading. The key idea is to map threads to output channels and iterate along pixel, instead of mapping threads to pixels and iterate along output channels, as illustrated in Algorithm \ref{ILP-M}. Some secondary well-known optimization strategy, such as transposing the output images for coalescing write, are ignored for simplification.

\begin{figure}[h!]
\centering
\captionsetup[subfloat]{farskip=2pt,captionskip=1pt}

\subfloat[Direct Convolution]{\includegraphics[width=0.45\linewidth]{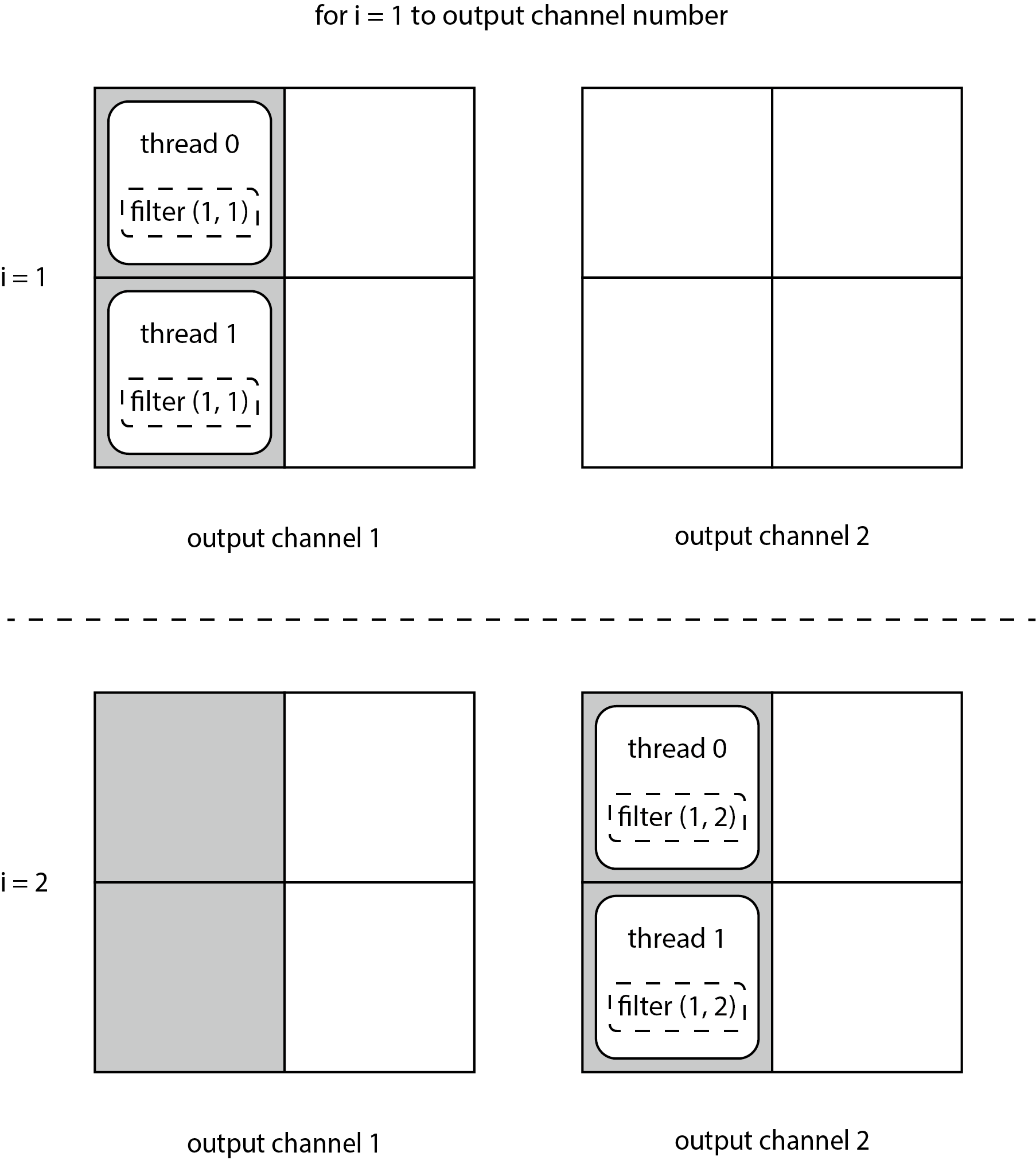}}
\hspace{1.2cm}
\subfloat[ILP-M Convolution]{\includegraphics[width=0.45\linewidth]{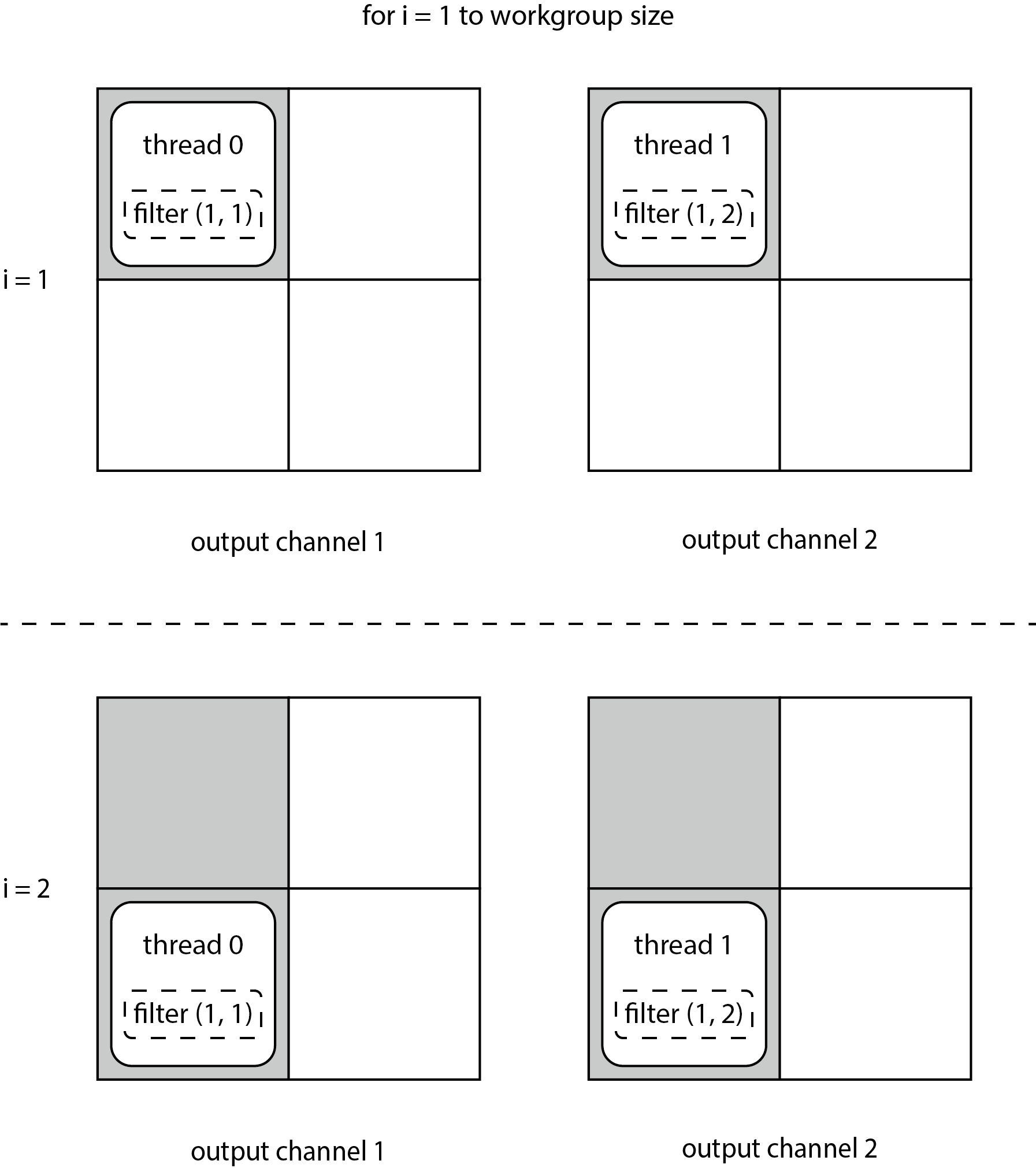}}

\caption{Difference of Direct Convolution and ILP-M Convolution. $filter(i, j)$ refers to the convolution filter of input channel $i$ and output channel $j$. The gray square indicates this pixel has been calculated. \label{ccoc}}

\end{figure}

For ILP-M convolution, all threads in the same workgroup also copy input images from the global memory into the shared memory collaboratively, and therefore a memory barrier is needed before accessing the input images in the shared memory. In this stage, each thread is mapped to a pixel. After that, threads of a workgroup are mapped to output channels rather than pixels. Each thread calculates the whole output image tile of its corresponding output channel, while for direct convolution, each thread calculates its corresponding pixel of all output channels. As all pixels of the same output channel is calculated with the same convolution filter, each thread loads and only needs to load one convolution filter (\texttt{filter\_size} values) for \texttt{workgroup\_size} arithmetic. Figure \ref{ccoc} illustrates the difference of convolution filters footprint between ILP-M convolution and direct convolution. The ratio of arithmetic instructions to global memory instructions is \texttt{workgroup\_size}, providing substantial space for compiles to reorder the instructions to hide the latency.

ILP-M convolution algorithm further reduces the register usage by iterating the weights of the convolution filter in the outer loop. Each time, the thread only needs to load one weight of the convolution filter, multiples it with all pixels of the input image tile and accumulates the result into the output image tile. After calculating the dot-product between all input channels and their corresponding convolution filters, the output images tile is written back to the global memory. As the threads compute a tile of the output images of different output channels, this global memory write is not coalesced due to the data layout of the output images. ILP-M convolution allows to chose whether to use shared memory to transpose the output images tiles and therefore, the output images tiles can be coalesced written back to global memory.



\begin{algorithm}
\caption{ILP-M Convolution Algorithm \label{ILP-M}}
\begin{algorithmic}[1]

\Variables
 \State LOCAL\_DIM\_\#, the shape of the workgroup at dim\#
 \State C, R, S, K, the input channel, filter width, filter height, output channel
\EndVariables
\State
\Function{convolution}{input, filter, output}
    
    
    \For{$in\_channel \gets 1$ to input\_channels}
        
        \State 
        
        \State \Call{load}{$img\_global$, $img\_shared$}
        \State \Call{barrier}{LOCAL\_MEM}
        \State 
        
 
        \For{$r \gets 1$ to FILTER\_WIDTH}  
            \For{$s \gets 1$ to FILTER\_WIDTH}
                \State filter\_reg $\gets$ filter[i][r][s][local\_id] \Comment{the filter is reorganized by [C][R][S][K] for coalescing read}
                \For{$wy \gets 1$ to LOCAL\_DIM\_Y}  
                    \For{$wx \gets 1$ to LOCAL\_DIM\_X}
                        \State out\_reg[wy][wx] += filter\_reg * $img\_shared$[wy + r][wx + s]
                    \EndFor
                \EndFor
            \EndFor
        \EndFor
        \State
    
    \EndFor
    \State
    
    \For{$wy \gets 1$ to LOCAL\_DIM\_Y}  
        \For{$wx \gets 1$ to LOCAL\_DIM\_X}
            \State save $out\_reg[wy][wx]$ to global memory
        \EndFor
    \EndFor

\EndFunction

\end{algorithmic}
\end{algorithm}

ILP-M convolution algorithm inherits all advantages of direct convolution in terms of the shared memory usage and arithmetic operations. As it is designed for single-image CNNs inference, ILP-M convolution hides the latency mainly by instruction-level parallelism rather than thread-level parallelism. ILP-M convolution algorithm maximizes the instruction-level parallelism by eliminating the instruction dependency and reducing the register usage. Meanwhile, the ratio of arithmetic instructions to memory instructions is quite high, providing enough space for compilers to fuse these two kinds of instructions to hide the latency.

\section{Evaluation and Experiments}

We implemented ILP-M convolution kernels\footnote{The code is available at: https://github.com/jizhuoran/sj\_convolution} with OpenCL for portability to embedded GPUs and integrated GPUs. We also implemented an auto-tuning library to chose the optimal combination of the kernel parameters, such as the tile size and workload per thread. We compared the five GPUs convolution algorithms discussed in previous sections, which are \texttt{im2col} convolution, \texttt{libdnn} convolution, Winograd convolution, direct convolution, and our ILP-M convolution. Other convolution kernels are also written in OpenCL, and we used the GEMM function of clBLAS as the matrix multiplication kernels.

The experiments were conducted on three typical and distinctive platforms: mobile GPUs (Arm Mali-G76 MP10), integrated GPUs (AMD Radeon Vega 8), and high-end dedicated GPUs (AMD Radeon VII). The details of the model type and configuration are shown in Table \ref{platform}. The high-end dedicated GPUs usually have many compute-units and dedicated graphics memory with extremely high bandwidth. Integrated GPUs share host memory with CPUs, whose bandwidth is quite limited. The compute-units of integrated GPUs are usually the same as the dedicated GPUs, while the number of compute-units is usually less than the high-end dedicated GPUs. For embedded GPUs or mobile GPUs, the memory bandwidth is similar to the integrated GPUs, but the compute-unit is simple and has less arithmetic logic units.


\begin{table*}[h]
\centering
\caption{Experiment Devices Configuration\label{platform}}
\begin{tabular}{l|l|l|l|l|l}
\hline
Model              & Global Memory Type  & Memory Bandwidth & CU & ALUs / CU & Total ALUs \\ \hline
AMD Radeon™ VII    & HBM2                & 1024 GB/s        & 60 & 64        & 3840       \\
AMD Radeon™ Vega 8 & DDR4 single-channel & 25 GB/s          & 8  & 64        & 512        \\
Arm Mali-G76 MP10       & LPDDR4 dual-channel & 33.3 GB/s        & 10 & 24        & 240        \\ \hline
\end{tabular}
\end{table*}

We ran the execution speed experiments with ResNet on ImageNet, which is the most popular and state-of-the-art CNNs architecture. There are five typical ResNet architectures, which has the same convolution layers. The difference is only the number of different convolution layers. All non-1x1 convolution layers of ResNet has three convolution filters (Table \ref{reset_setting}), except the first one that deals with the raw images, whose convolution filters are seven. Our experiments only cover these $three \times three$ convolution filter, as they are the central part of the ResNet.

\begin{table*}[h!]
\centering
\caption{Convolution Layers of ResNet\label{reset_setting}}
\begin{tabular}{l|l|l|l|l|l|l|l}
\hline
Layer   & C $\times$ K     & H $\times$ W   & ResNet 18 & ResNet 34 & ResNet 50 & ResNet 101 & ResNet 152 \\ \hline
conv2.x & 64 $\times$ 64   & 56 $\times$ 56 & 2 $\times$ 2     & 2 $\times$ 3     & 1 $\times$ 3     & 1 $\times$ 3      & 1 $\times$ 3      \\
conv3.x & 128 $\times$ 128 & 28 $\times$ 28 & 2 $\times$ 2     & 2 $\times$ 4     & 1 $\times$ 4     & 1 $\times$ 4      & 1 $\times$ 8      \\
conv4.x & 256 $\times$ 256 & 14 $\times$ 14 & 2 $\times$ 2     & 2 $\times$ 6     & 1 $\times$ 6     & 1 $\times$ 23     & 1 $\times$ 36     \\
conv5.x & 512 $\times$ 512 & 7 $\times$ 7   & 2 $\times$ 2     & 2 $\times$ 4     & 1 $\times$ 3     & 1 $\times$ 3      & 1 $\times$ 3      \\ \hline
\end{tabular}
\end{table*}


\subsection{Result}

\begin{figure*}[h!]
\centering
\captionsetup[subfloat]{farskip=2pt,captionskip=1pt}

\subfloat{\includegraphics[width=0.24\linewidth]{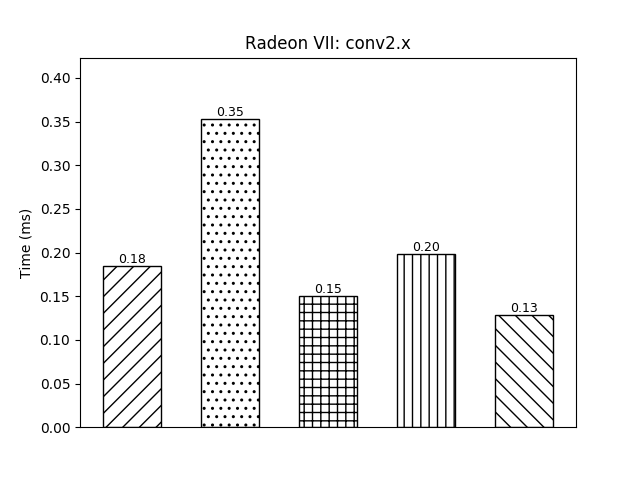}}
\subfloat{\includegraphics[width=0.24\linewidth]{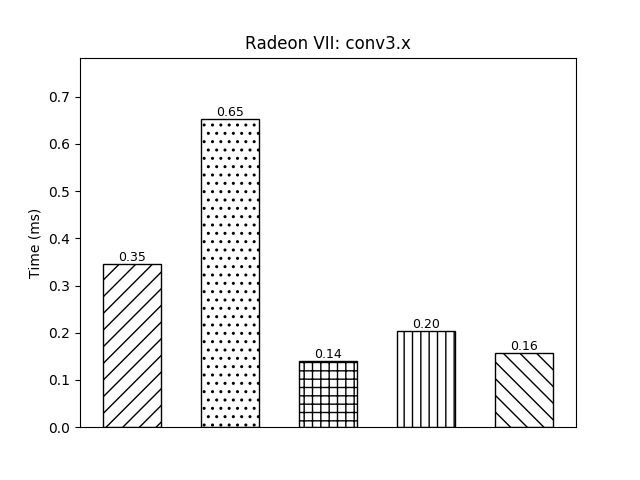}}
\subfloat{\includegraphics[width=0.24\linewidth]{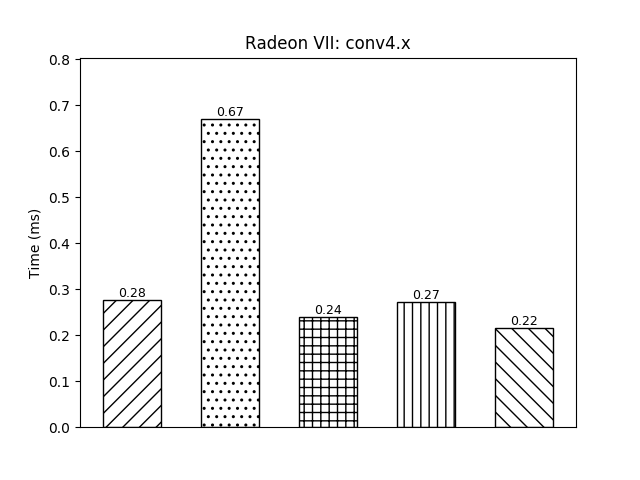}}
\subfloat{\includegraphics[width=0.24\linewidth]{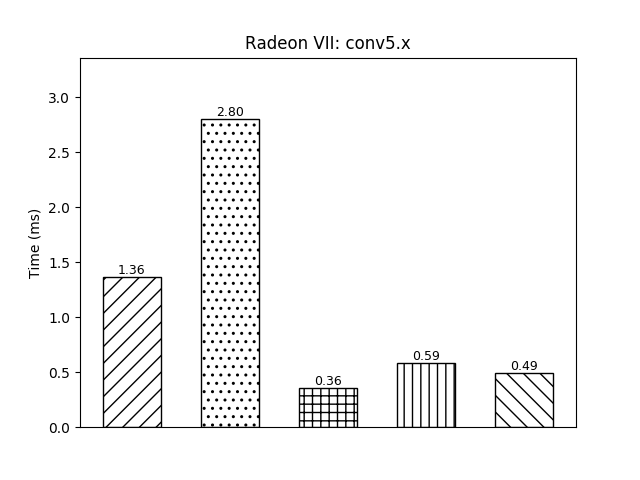}}\\

\subfloat{\includegraphics[width=0.24\linewidth]{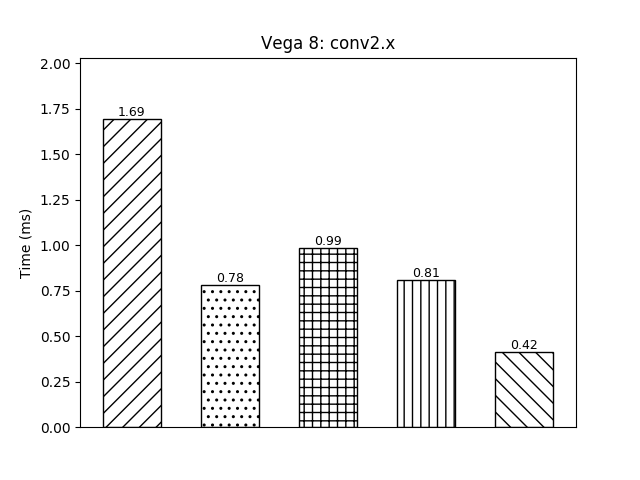}}
\subfloat{\includegraphics[width=0.24\linewidth]{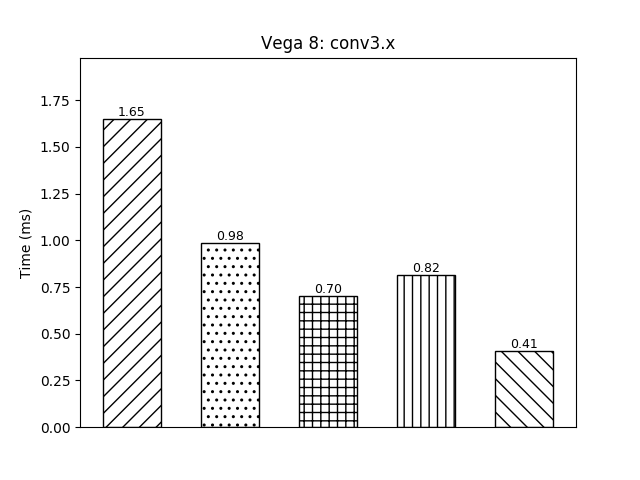}}
\subfloat{\includegraphics[width=0.24\linewidth]{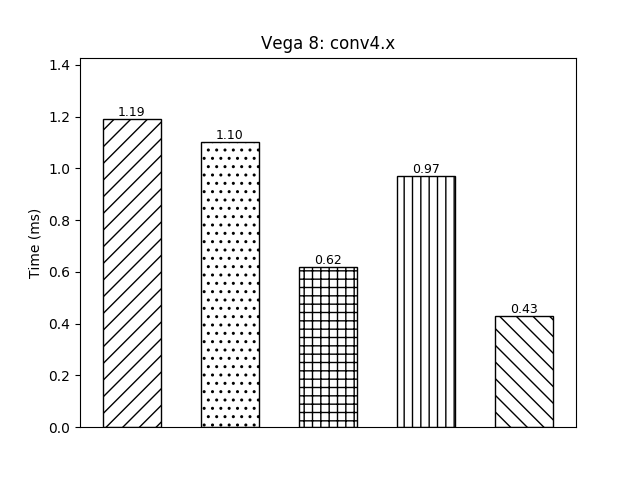}}
\subfloat{\includegraphics[width=0.24\linewidth]{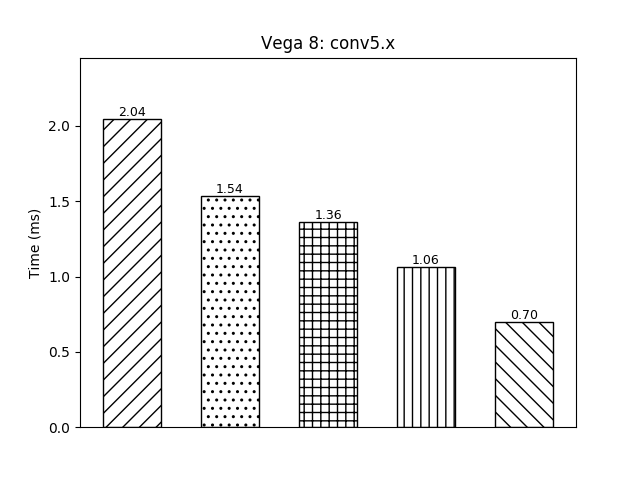}}\\

\subfloat{\includegraphics[width=0.24\linewidth]{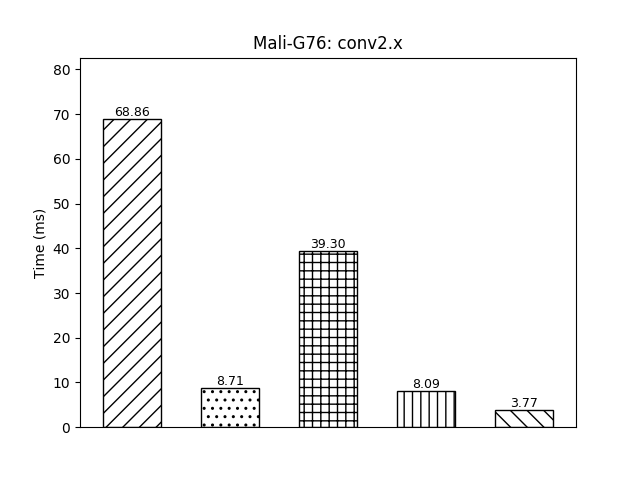}}
\subfloat{\includegraphics[width=0.24\linewidth]{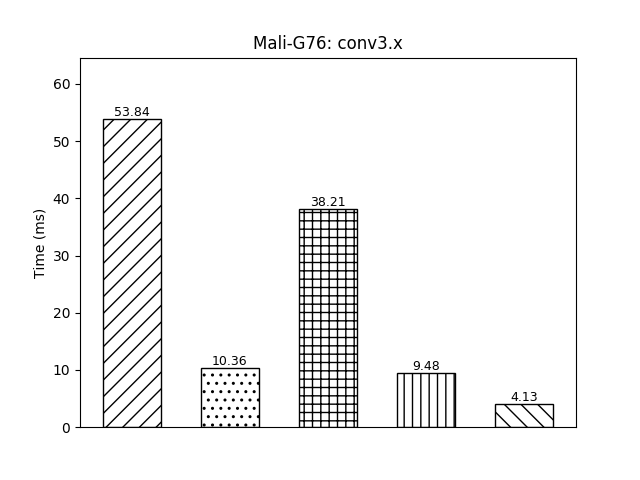}}
\subfloat{\includegraphics[width=0.24\linewidth]{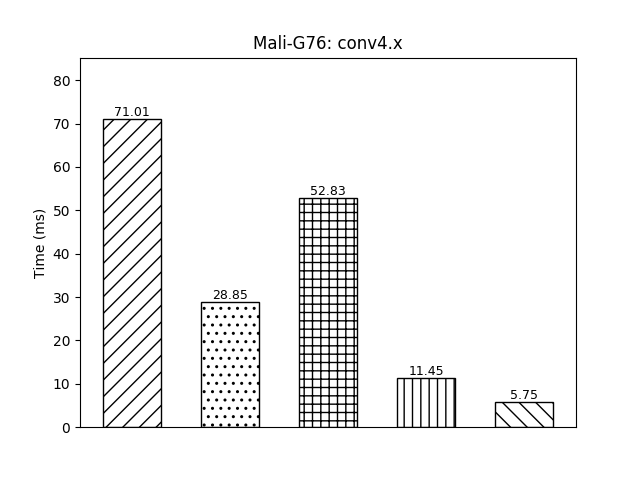}}
\subfloat{\includegraphics[width=0.24\linewidth]{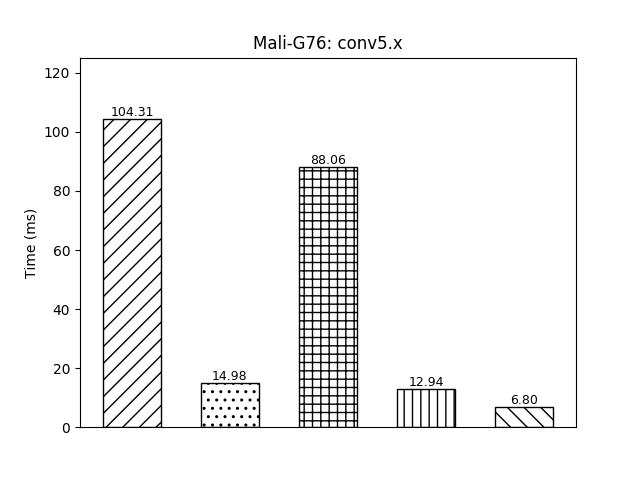}} \\

\subfloat{\includegraphics[width=0.3\linewidth]{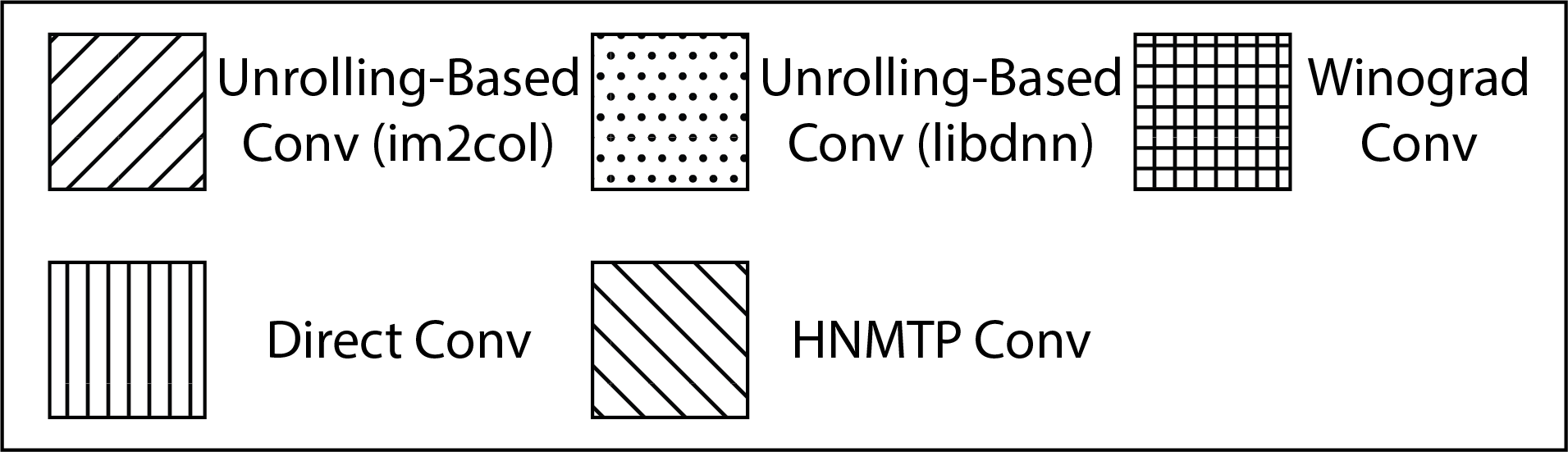}}

\caption{Comparison of the Execution Time \label{result}}

\end{figure*}

Figure \ref{result} shows the CNNs inference time of different convolution layers on different kinds of GPUs. The ILP-M convolution algorithm surpasses all other convolution algorithms in all convolution layers on embedded mobiles and integrated GPUs. On dedicated GPUs, the fastest CNNs inference is achieved by either ILP-M convolution or Winograd convolution. The thread-level parallelism is so insufficient with only one image that the convolution algorithm with highest instruction-level parallelism has the best performance.

As the \texttt{libdnn} convolution algorithm eliminates the global memory access overhead of the unrolled input matrix, it overperforms the \texttt{im2col} convolution algorithm on both integrated GPUs and mobile GPUs, whose memory bandwidth is quite limited. However, on dedicated GPUs, \texttt{libdnn} convolution is more than $2 \times$ slower than \texttt{im2col} convolution, as the memory overhead is negligible with such high global memory bandwidth. It confirms our knowledge that most deep learning frameworks use \texttt{im2col} convolution algorithm for both CNNs training and testing.

The execution time of Winograd convolution algorithm is short than that of \texttt{im2col} convolution algorithm in all cases. The Winograd convolution has the fewest floating-point multiplication complexity among all algorithms. The high thread-level parallelism of Winograd convolution results in up to $3.78 \times$ speedup compared with the \texttt{im2col} convolution on dedicated GPUs, which has many compute-units. However, both the Winograd convolution and \texttt{im2col} convolution perform poorly on Mali-G72. Both of them rely on GEMM a lot, which needs large workgroup to reduce global memory access. However, as the compute-units of Mali-G72 has fewer ALUs than the other two, it favors a smaller workgroup size.

As direct convolution has too many parameters and optimization strategies, the performance varies a lot among different platforms.  On integrated GPUs, the performance of direct convolution is slightly better than the \texttt{libdnn} convolution in most cases, while direct convolution is usually slower than Winograd convolution. In contrast, the execution time of direct convolution is short than that of \texttt{libdnn} convolution on embedded GPUs. The direct convolution has around the same global memory access with \texttt{libdnn} convolution but needs less index and memory address calculations. With the saving of arithmetic operations, direct convolution overwhelms the \texttt{libdnn} on the less powerful embedded GPUs. However, direct convolution needs lots of efforts to tune the kernels to find out the optimal combination of parameters.

In our experiments, ILP-M convolution overwhelms all existing convolution algorithms on integrated GPUs and embedded GPUs. On integrated GPUs, ILP-M convolution reduces the execution by up to $46.8\%$, compared with the second-fastest convolution algorithm, Winograd convolution. While Winograd convolution reduces arithmetic operations by additional global memory access, ILP-M convolution regards global memory access as more expensive operations. As the memory bandwidth of the GPUs without dedicated graphics memory is quite limited, even moderate global memory access may become the bottleneck. On the other hand, direct convolution is the fastest existing convolution algorithm on embedded GPUs, and our ILP-M convolution achieves $2.30 \times$ speedup with fewer efforts for GPU kernels tuning. ILP-M convolution is improved from the direct convolution algorithm and inherits all its advantages. ILP-M convolution has much higher instruction-level parallelism, as it has more independent memory instructions and arithmetic instructions to fuse and fewer register usage.

\subsection{Profile}

In this section, we analyzed the performance of different convolution algorithm in the granularity of kernel-level with the profiling result in terms of memory and arithmetic instructions. We chose the most frequent \texttt{conv4.x} as the example, which has $256$ input channels and output channels and $14 \times 14$ pixels per image. We conducted the run-time profile on Vega 8 with codeXL, which provides thoughtful profiling information. The \texttt{im2col} convolution consists of the \texttt{im2col} kernel and GEMM kernel, while the Winograd convolution consists of a \texttt{trans\_from\_images} kernel, $16$ GEMM kernels, and a \texttt{trans\_to\_output} kernel. The GPU kernel that transforms the filters is ignored as the filters are constant in CNNs inference, which can be computed offline.

\subsubsection{Memory Metrics}

\begin{table*}[h!]
\scalebox{0.9}{
\begin{tabular}{l|p{2cm}|p{2cm}|p{2cm}|p{3cm}|p{2cm}}
\hline
Kernel(s)                    & Global Memory Read (MB) & Global Memory Write (MB) & Memory Unit Busy (\%) & Shared Memory Usage (Btye/Workgroup) & Shared Memory Bank Conflit (\%) \\ \hline
im2col\_im2col               & 0.20            & 1.73             & 48.91            & 0                   & 0                          \\
im2col\_gemm                 & 9.27            & 0.20             & 24.45            & 4224                & 0                          \\
libdnn\_conv                 & 2.48            & 0.20             & 15.19            & 4480                & 0.34                       \\
winograd\_trans\_from\_image & 0.20            & 0.77             & 25.01            & 1408                & 0.36                       \\
winograd\_gemm (16 times)    & 4.91            & 0.77             & 13.49            & 4224                & 0                          \\
winograd\_trans\_to\_output  & 0.77            & 0.19             & 69.96            & 0                   & 0                          \\
direct\_conv                 & 2.60            & 0.19             & 81.29            & 512                 & 4.27                      \\
ILP-M\_conv                   & 2.46            & 0.20             & 14.84            & 1024                 & 0                          \\ \hline
\end{tabular}
}
\caption{Profile Metrics Related to Memory \label{pmrm}}
\end{table*}

As shown in Table \ref{pmrm}, ILP-M convolution is one of the convolution algorithms that have the least number of global memory access. It reduces the global memory read by $74.0\%$ and $58.2\%$ and global memory write by $89.6\%$ and $88.4\%$ compared with the \texttt{im2col} convolution and Winograd convolution. With the help of L2 cache, direct convolution has similar global memory access numbers with ILP-M convolution but the memory units busy time is much higher than that of ILP-M convolution due to the duplicated convolution filters loading. 

Meanwhile, ILP-M convolution needs the least size of the shared memory per thread. Both ILP-M convolution and direct convolution cache the input images only, while other convolution algorithms also need to cache the convolution filters. However, the number of warps is usually insufficiency for single-image CNNs inference, so the usage of shared memory is usually not the bottleneck.

ILP-M convolution has no shared memory band conflicts of its main computation kernels. The threads with the same warp read the same data from the shared memory at each time. Thanks to the broadcast mechanism, only one shared memory access is needed. In contrast, \texttt{libdnn} convolution and direct convolution may need to access different data of the same shared memory bank, incurring shared memory band conflicts. The serialized shared memory accesses reduce performance and waste energy.

\subsubsection{Arithmetic Metrics}
\begin{table*}[]
\centering
\scalebox{0.9}{
\begin{tabular}{l|l|l|l|l}
\hline
Kernel(s)                    & Wavefronts & Total Vector Inst ($10^4$) & Total Scalar Inst ($10^4$) & Vector ALU Busy (\%) \\ \hline
im2col\_im2col               & 784        & 248.32                     & 343.68                     & 10.09                \\
im2col\_gemm                 & 224        & 4707.2                     & 785.76                     & 44.31                \\
libdnn\_conv                 & 64         & 6289.12                    & 1277.28                    & 45.73                \\
winograd\_trans\_from\_image & 256        & 112.16                     & 27.84                      & 10.04                \\
winograd\_gemm               & 1024       & 2469.12                    & 447.36                     & 41.24                \\
winograd\_trans\_to\_output  & 256        & 52.8                       & 2.88                       & 7.21                 \\
direct\_conv                 & 256        & 5711.52                    & 990.88                     & 31.47                \\
ILP-M\_conv                  & 32         & 3935.2                     & 43.84                      & 55.86                \\ \hline

\end{tabular}
}
\caption{Profile Metrics Related to Arithmetic \label{pmra}}
\end{table*}

ILP-M convolution uses the second least number of instructions (Table \ref{pmra}), which includes vector instructions and scalar instructions. The number of instructions issued by ILP-M convolution is only $65.4\%$ of \texttt{im2col} convolution, $52.6\%$ of \texttt{libdnn} convolution, and $59.4\%$ of the direct convolution. Even though ILP-M performs the same number of useful floating-point arithmetic, it needs less global memory address calculation and vector accesses than these convolution algorithms. As \texttt{libdnn} convolution needs to unroll the same image tile multiple times, it needs the most vector instructions due to the redundant memory address calculations.

The total number of instructions of ILP-M convolution is $1.29 \times$ of the Winograd convolution. However, there are fewer memory barriers in ILP-M convolution, which improves the instruction-level parallelism. Between two barriers, ILP-M convolution has both arithmetic instructions and memory instructions, while GEMM kernels of Winograd only have arithmetic instructions. The compilers can fuse these two kinds of instructions to hide the latency and further improves the instruction-level parallelism. As the vector ALUs are not always busy in single-image CNNs inference, high instruction-level parallelism means high vector ALUs busy time, and therefore more vector instructions can be executed in a unit time. In other words, the ILP-M convolution needs less time to execute its vector instructions, even if it has more vector instructions.

Direct convolution has the same number of memory barriers with the ILP-M convolution. However, its instruction-level parallelism is far less than ILP-M convolution. To pipeline the same number of calculations, say $N$ calculations ($N < $ \texttt{workgroup\_size}), direct convolution needs $N$ weights of convolution filters, which needs $N$ register to store as they are loaded from the global memory. In contrast, ILP-M convolution only needs one register to store one weight. With the same register usage constraints, the compiler can pipeline more calculations for ILP-M convolution.

\section{Conclusion and Future Work}

In this report presents the fastest convolution algorithm for single-image convolution neutral network inference on mobile GPUs. In the future, we are going to supports more tuning options, such as workload per threads and output coalescing write.

\bibliographystyle{unsrt}  
\bibliography{hnmtp_conv}  

\begin{thebibliography}{1}

\bibitem{im2col}
Xu~Han.
\newblock Implement convolution in cnn.
\newblock {\em http://xuhan.me/2016/09/09/conv/}, 2016.

\bibitem{tschopp2016efficient}
Fabian Tschopp, Julien~NP Martel, Srinivas~C Turaga, Matthew Cook, and Jan
  Funke.
\newblock Efficient convolutional neural networks for pixelwise classification
  on heterogeneous hardware systems.
\newblock In {\em 2016 IEEE 13th International Symposium on Biomedical Imaging
  (ISBI)}, pages 1225--1228. IEEE, 2016.

\bibitem{Tschoppgithub}
Tschopp.
\newblock Opencl caffe.
\newblock \url{https://github.com/BVLC/caffe/tree/opencl}, 2018.

\bibitem{lavin2016fast}
Andrew Lavin and Scott Gray.
\newblock Fast algorithms for convolutional neural networks.
\newblock In {\em Proceedings of the IEEE Conference on Computer Vision and
  Pattern Recognition}, pages 4013--4021, 2016.

\bibitem{krizhevsky2012cuda}
Alex Krizhevsky.
\newblock cuda-convnet: High-performance c++/cuda implementation of
  convolutional neural networks.
\newblock {\em Source code available at https://github.
  com/akrizhevsky/cuda-convnet2 [March, 2017]}, 7, 2012.

\bibitem{perrot2016optimized}
Gilles Perrot, St{\'e}phane Domas, and Rapha{\"e}l Couturier.
\newblock An optimized gpu-based 2d convolution implementation.
\newblock {\em Concurrency and Computation: Practice and Experience},
  28(16):4291--4304, 2016.

\bibitem{iandola2013communication}
Forrest~N Iandola, David Sheffield, Michael~J Anderson, Phitchaya~Mangpo
  Phothilimthana, and Kurt Keutzer.
\newblock Communication-minimizing 2d convolution in gpu registers.
\newblock In {\em 2013 IEEE International Conference on Image Processing},
  pages 2116--2120. IEEE, 2013.

\bibitem{chen2017optimizing}
Xiaoming Chen, Jianxu Chen, Danny~Z Chen, and Xiaobo~Sharon Hu.
\newblock Optimizing memory efficiency for convolution kernels on kepler gpus.
\newblock In {\em 2017 54th ACM/EDAC/IEEE Design Automation Conference (DAC)},
  pages 1--6. IEEE, 2017.

\bibitem{li2015fast}
Shigang Li, Yunquan Zhang, Chunyang Xiang, and Lei Shi.
\newblock Fast convolution operations on many-core architectures.
\newblock In {\em 2015 IEEE 17th International Conference on High Performance
  Computing and Communications, 2015 IEEE 7th International Symposium on
  Cyberspace Safety and Security, and 2015 IEEE 12th International Conference
  on Embedded Software and Systems}, pages 316--323. IEEE, 2015.

\end{thebibliography}

\end{document}